\documentclass[iop]{emulateapj}
\usepackage{mathrsfs}

\begin{document}

\title{Crystal chemistry of three-component white dwarfs and neutron star crusts: \\phase stability, phase stratification, and physical properties}

\shorttitle{Crystal chemistry of white dwarfs and neutron star crusts} 

\author{T. A. Engstrom, N. C. Yoder, V. H. Crespi}
\affil{Department of Physics, The Pennsylvania State University, University Park, PA 16802, USA}
\email{tae146@psu.edu\\ncy5007@psu.edu\\vhc2@psu.edu}

\begin{abstract}
A systematic search for multicomponent crystal structures is carried out for five different ternary systems of nuclei in a polarizable background of electrons, representative of accreted neutron star crusts and some white dwarfs.  Candidate structures are ``bred'' by a genetic algorithm, and optimized at constant pressure under the assumption of linear response (Thomas-Fermi) charge screening.  Subsequent phase equilibria calculations reveal eight distinct crystal structures in the $T=0$ bulk phase diagrams, five of which are complicated multinary structures not before predicted in the context of compact object astrophysics.  Frequent instances of geometrically similar but compositionally distinct phases give insight into structural preferences of systems with pairwise Yukawa interactions, including and extending to the regime of low density colloidal suspensions made in a laboratory.  As an application of these main results, we self-consistently couple the phase stability problem to the equations for a self-gravitating, hydrostatically stable white dwarf, with fixed overall composition.  To our knowledge, this is the first attempt to incorporate complex multinary phases into the equilibrium phase layering diagram and mass-radius-composition dependence, both of which are reported for He-C-O and C-O-Ne white dwarfs.  Finite thickness interfacial phases (``interphases'') show up at the boundaries between single-component bcc crystalline regions, some of which have lower lattice symmetry than cubic.  A second application -- quasi-static settling of heavy nuclei in white dwarfs -- builds on our equilibrium phase layering method.  Tests of this nonequilibrium method reveal extra phases which play the role of transient host phases for the settling species.  
\end{abstract}


\section{Introduction}

When an impure white dwarf (WD) or neutron star crust (NSC) is slowly cooled from above its melting temperature, one expects the extra compositional degrees of freedom are taken advantage of to form crystals which are more efficiently packed than phase-separated bcc lattices.  Indeed, several investigators have considered non-Bravais and multicomponent lattices as the possible ground state of astrophysical compact objects.  One of the earliest was \citet{dys71}, who suggested a rock salt structure of Fe and He nuclei might be stable.  More recently, \citet{koz12} studied cesium-chloride and magnesium-diboride structures within the Coulomb crystal model.  \citet{kob14} have argued that the ground state structure above neutron drip density may be similar to that of the displacive ferroelectric BaTiO$_3$, due to the symmetry-lowering effect of interstitial neutrons on a bcc lattice of nuclei.  A related line of inquiry concerns the freezing of multicomponent ion plasmas from the liquid state.  See \cite{med10} for a semi-analytic calculation and references to earlier numerical methods. One such method -- classical molecular dynamics -- has been used extensively to simulate a multicomponent plasma with the \citet{gup07} composition \citep{hor07, hor09pre, hor09prc}.  The latter of these works features a 14-component, $\approx\,$28,000 particle system which was annealed for $\sim\,$$10^7$ phonon cycles below the melting temperature.  A dominantly Se ($Z$=34) bcc lattice was formed, with small-$Z$ nuclei occupying interstitial positions and larger-$Z$ nuclei acting as substitutional impurities.  In addition, there was a tendency for small-$Z$ nuclei to cluster together, forming an effective large-$Z$ particle.  In a different simulation where annealing was again carried out for $\sim\,$$10^7$ phonon cycles \citep{hor09pre}, phase-separated regions (microcrystals) formed in the solid phase.  One phase was depleted in small-$Z$ nuclei, while another was enriched.

Simulated annealing is an excellent means for directly modeling the dynamics of crystalline systems, but it often cannot access the very long timescales associated with the nucleation and growth of complex multicomponent crystal phases, due to the exponentially slow dynamics of surmounting reaction barriers against the complex cooperative rearrangements needed to form such crystals. For example, terrestrial carbon steels, which typically have only 2--3 alloying elements, must be annealed for a minimum of $\sim\,$$10^{13}$ phonon cycles ($\sim\,$10 seconds) to find their ground state \citep{asm77}.  Alternative methods including random structure searching \citep{pic11}, particle swarm optimization \citep{wan10}, and genetic/evolutionary search techniques \citep{oga06, abr08, wu14} have been applied with great success to this ``crystal structure problem," but have not yet been applied at the extreme conditions of compact astrophysical objects.  When coupled with an appropriate description of the (fully pressure-ionized) microphysics, such methods could provide a means to efficiently search for new crystal structures in multicomponent WDs and NSCs, complementing the existing simulated annealing work.  

The existence of lower-symmetry (i.e. non-cubic) and/or multinary phases within WDs and NSCs could have several astrophysical implications.  Most astrophysical calculations assume the material is a bcc polycrystal with grain sizes small compared to the other macroscopic physical scales in the problem. Therefore, for example, the rank-four elastic tensor is averaged and smoothed to produce a scalar shear modulus relating the strain response to an applied stress (one popular averaging procedure is described by \citet{oga90}). The possibility of multiple, complicated lattice structures, and preferential alignment with e.g. the local magnetic field, would necessitate computing the full elastic tensor.  Anisotropies, soft phonon modes, and elastic instabilities such as the incipient ones described in \citet{eng15} could have significant effects on elasticity-related astrophysical observables such as magnetar flares \citep{per11}, related quasi-periodic oscillations \citep{isr05}, and possibly some pulsar glitches \citep{cha08}. It could also significantly affect the future observability of gravitational-wave emission, both in the context of magnetar flares and continuous waves \citep{joh13}. Grain/phase domain boundaries would lead to preferred stress-failure locations, and on large scales might affect dissipation of modes involving the crust such as torsion or shear modes (\citet{isr05}, used to explain quasi-periodic oscillations after magnetar flares) and $r$-modes (similar to the ``crust freezing" scenario in \citet{lin00}). These again would have implications both for electromagnetic and gravitational wave observations.   Another kind of implication has to do with the composition of WD debris disks and planetary systems, inferred from metal abundances in the accreting WD's atmosphere \citep{bar12, raf11}.  Entering into this calculation is the settling rate of the high-$Z$ metals. In principle, this rate depends on the buoyancy of the settling species' host phase(s) as well as the microphysics involved in ordinary grain growth processes, namely interfacial energies and grain boundary mobilities \citep{kri02}.  

In this work we carry out a systematic search for the ground state crystal structure of three-component systems at conditions relevant to WDs and NSCs.  The main goals are 1) to identify possible new phases through global search of the multicomponent crystal structure phase diagram, and 2) to determine in what contexts those phases might appear in WDs and NSCs through sample applications to layering stability.  To the first end, we employ a popular genetic search algorithm.  The lowest enthalpy structures found by the genetic search are included in bulk phase diagram calculations, which reveal five new complicated binary and ternary crystal structures, four having sub-cubic lattice symmetry.  To the second end, we demonstrate a self-consistent coupling of the phase stability calculation with the basic equations of a self-gravitating, hydrostatically stable white dwarf. Several compositional instances of the newly found binary phases show up as finite thickness ``interphases" between pure bcc strata in cold, He-C-O and C-O-Ne white dwarfs.  Additional binary phases make a transient appearance in nonequilibrium settling calculations, as host phases for the settling species. 

\section{bulk phase diagram calculation}

This section describes a global search of composition and structure, using five ternary systems of nuclei thought to be relatively prevalent in WD or NSC matter, and covering a range of distinct ``crystal chemistries."   The starting point is an effective Hamiltonian for completely pressure-ionized matter.  We work within linear response theory -- see Section V of \citet{pol73}, and \citet{bai02}, for example.  In this framework, a system of point nuclei (with charges $Z_ie$ and static positions $\mathbf{r}_i$) immersed in a polarizable, charge compensating background of electrons has kinetic plus electrostatic potential energy
\begin{eqnarray}
E = T_0 + \frac{e^2}{2}&\Bigg{\{}& \sum_{i\neq j}Z_iZ_j\int\frac{d^3k}{(2\pi)^3}\frac{4\pi e^{i\mathbf{k}\cdot(\mathbf{r}_i-\mathbf{r}_j)}}{k^2\epsilon(\mathbf{k})} \label{E}\\ 
&+& \sum_iZ_i^2\int\frac{d^3k}{(2\pi)^3}\frac{4\pi}{k^2}\bigg[\frac{1}{\epsilon(\mathbf{k})}-1\bigg] \nonumber\\ 
&-& \sum_{i,\,j}Z_iZ_j\;\frac{1}{V}\int d^3r\int\frac{d^3k}{(2\pi)^3}\frac{4\pi e^{i\mathbf{k\cdot r}}}{k^2\epsilon(\mathbf{k})} \Bigg{\}},\nonumber 
\end{eqnarray}
(compare \citet{bai02} Equations 1-3).  It is not immediately obvious that the above Hamiltonian includes the leading order correction to the kinetic energy $T_0$ of the uniform electron gas (it does).  This can be seen by expanding the kinetic energy in powers of the density nonuniformity correction: $T=T_0 + \frac{e^2}{2}\int d^3r\, d^3r'\, \delta n_e(\mathbf{r})G(\mathbf{r-r'})\delta n_e(\mathbf{r}') + \dots$ and keeping only the first two terms such that a total energy minimization identifies $-G(\mathbf{k})^{-1}$ as the static response function of the uniform gas. Equation \ref{E} also contains all Coulomb interactions except for the infinite nuclear self energies.  With a choice of the simple Thomas-Fermi dielectric function $\epsilon_{TF}(\mathbf{k})=1+k_0^2/k^2$, the integrals are standard ones and the Hamiltonian reduces to
\begin{eqnarray}
E_{TF} = T_0 + \frac{e^2}{2}\Bigg{\{} &-&k_0\sum_iZ_i^2 - \frac{4\pi}{k_0^2V}\sum_{i,\,j}Z_iZ_j\nonumber\label{ETF}\\
&+& \sum_{i\neq j}\frac{Z_iZ_je^{-k_0|\mathbf{r}_i-\mathbf{r}_j|}}{|\mathbf{r}_i-\mathbf{r}_j|} \Bigg{\}}.
\end{eqnarray}
In performing structural optimizations, one converges the energy by working with a supercell of volume $V_c$ and $N-1$ periodic copies thereof.  If $\mathbf{R}$ is a primitive lattice vector (supercell translation vector) and $p,q$ index the basis, the total energy per supercell is written
\begin{eqnarray}
\frac{E_{TF}}{N} = \tau_0V_c + \frac{e^2}{2}\Bigg{\{} &-&k_0\sum_pZ_p^2 - \frac{4\pi}{k_0^2V_c}\sum_{p,\,q}Z_pZ_q\nonumber\label{ETF/N}\\
&+& {\sum_{\mathbf{R},\,p,\,q}}'\;\frac{Z_pZ_qe^{-k_0\mathscr{R}_{pq}}}{\mathscr{R}_{pq}} \Bigg{\}},
\end{eqnarray}
where $\mathscr{R}_{pq} = |\mathbf{R}+\mathbf{r}_p-\mathbf{r}_q|$ and the prime on the last sum indicates that terms with $\mathscr{R}_{pq}=0$ are excluded.  
\pagebreak
For this work we use kinetic energy density 
\begin{eqnarray}
\tau_0 &=& \frac{1}{8\pi^2}\frac{m_ec^2}{\lambda_e^3} \Big[ \frac{x^2(1+2x^2)}{\beta} - \ln\Big(x+\frac{x}{\beta}\Big)\Big] - m_ec^2n_e,\nonumber\label{tau0}\\
&&
\end{eqnarray}
and screening length
\begin{equation}
k_0^{-1} = \frac{\lambda_e}{2x}\sqrt{\frac{\pi\beta}{\alpha}}, \label{k0inverse}
\end{equation}
corresponding to the relativistic, degenerate gas.  Here $\lambda_e=\hbar/m_ec$ is the reduced Compton wavelength, $x=p_F/m_ec = \lambda_e(3\pi^2n_e)^{1/3}$, $\beta=x/\sqrt{1+x^2}$, and $\alpha\approx1/137$ is the fine structure constant.  The Thomas-Fermi description breaks down when the screening length localizes electrons to within their Compton wavelength; this occurs for $x\gtrsim10$ ($\rho\gtrsim10^9$ g/cc). Approaching this extreme relativistic limit, the ratio of screening length to Wigner-Seitz radius $r_s = (3\langle Z\rangle/4\pi n_e)^{1/3}$ tends to a constant (for a bcc lattice of Fe nuclei, the constant is 1.82), so we might anticipate that phase boundaries become stationary in this scale-invariant limit.  Saturation of $k_0^{-1}/r_s$ at this small value is indicative of the over-screening predicted by the Thomas-Fermi model.  Within these constraints however, the model has the advantage of being both reasonably accurate and computationally efficient, readily incorporated into a global search of crystal structure and composition.

Structural optimizations using Equations \ref{ETF/N}--\ref{k0inverse} are conveniently carried out at constant volume -- one need only minimize a pairwise sum of effective Yukawa interactions, which converges rapidly in real space for $k_0^{-1}/r_s\sim1$.  The simplicity of this method is conducive to a large basis, useful for studying interfaces, defects and failure mechanisms (see for example, \citet{hor09prl}).  Because it will prove useful for the phase layering calculation described later, we choose instead to perform structural optimizations at constant external pressure $P$ and minimize the enthalpy.  This can be accomplished using a modified version of the General Utility Lattice Program (GULP) \citep{gal03}, where the Yukawa interaction (available as a special case of the ``general" pair potential) is given the capability to handle a $V_c$-dependent screening length.  We modify GULP's enthalpy per unit cell to $h_{TF}=E_{TF}/N+PV_c$, and the first strain derivatives of the enthalpy (sufficient for steepest descents and conjugate gradient methods) are accordingly modified to
\begin{eqnarray}
\frac{\partial h_{TF}}{\partial \epsilon_{\alpha\beta}} & = & \Big[(P-P_0)V_c + \frac{2\pi e^2(2-\beta^2)}{3k_0^2V_c}\sum_{p,\,q}Z_pZ_q\\ 
&+& \frac{e^2k_0(1+\beta^2)}{12}\sum_{\mathbf{R},\,p,\,q}\;Z_pZ_qe^{-k_0\mathscr{R}_{pq}}\Big]\delta_{\alpha\beta} \nonumber \\
& - & \frac{e^2}{2}{\sum_{\mathbf{R},\,p,\,q}}'\;\frac{Z_pZ_q\,\mathscr{R}_{pq}^{\alpha}\,\mathscr{R}_{pq}^{\beta}\,e^{-k_0\mathscr{R}_{pq}}}{\mathscr{R}_{pq}^2}\Big(k_0 + \frac{1}{\mathscr{R}_{pq}}\Big),\nonumber
\end{eqnarray}  
where $P_0=n_e^2\partial (\tau_0/n_e)/\partial n_e$ is the kinetic pressure of the uniform electron gas.  Derivatives of $h_{TF}$ with respect to GULP's remaining degrees of freedom (fractional basis coordinates) are not affected by $k_0 \to k_0(V_c)$.

\begin{figure}[h]
\plotone{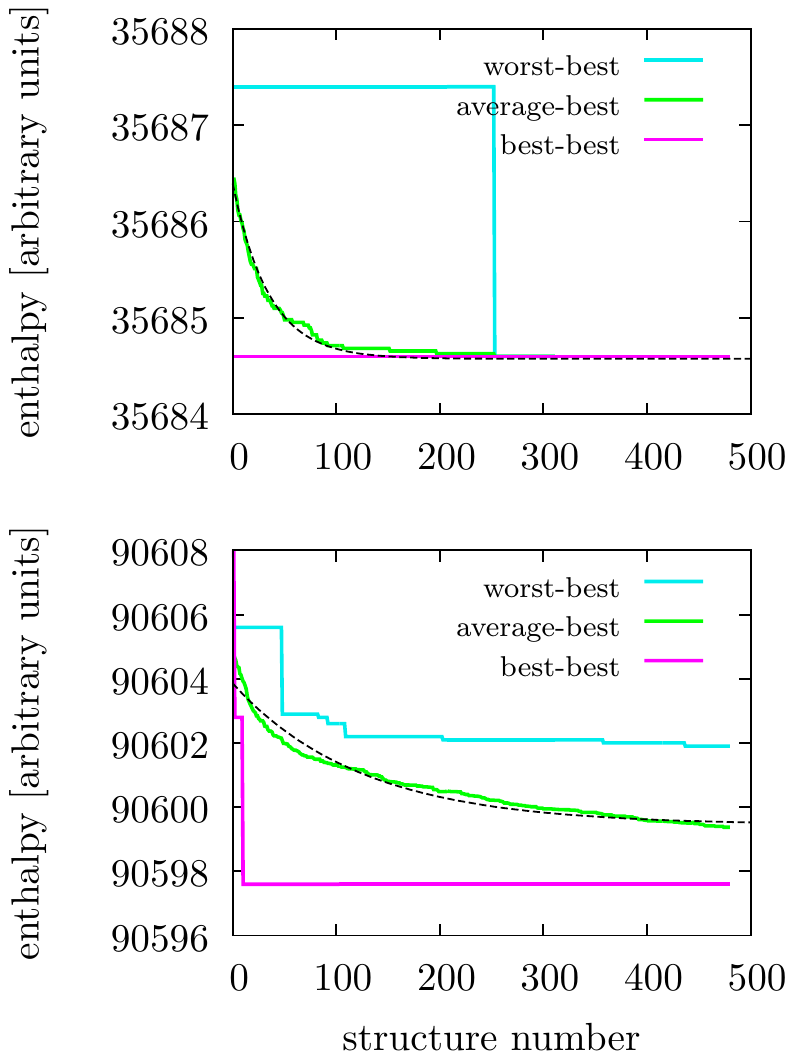}
\caption{  
Hartke plots for search cells FeO$_3$C$_2$ (top) and Fe$_4$O$_4$C$_4$ (bottom).  $y(x)$ is the enthalpy of the lowest enthalpy structure found within the range of structure numbers: zero to $x$. The initial 20 seed structures (randomly generated by XtalOpt) are not included, thus the Hartke plot contains information only about structures which have been genetically-operated-on.  For each plot, 100 runs were made with identical parameters. In the case where all runs eventually found the same lowest enthalpy structure (top), the worst-best is the single run that took the longest to find it.  In the case where not all runs found the same lowest enthalpy structure (bottom), the worst-best is the single run whose winning structure had the highest enthalpy.  Best-best was the quickest to find the overall lowest enthalpy structure, and average-best is the average over all 100 runs.  The Hartke lifetimes associated with the exponential fits to average-best (black dashed lines) are 35 (top) and 123 (bottom).  The winning structure found in the FeO$_3$C$_2$ search appears in the bulk phase diagram, as the $\eta$ phase.  The Fe$_4$O$_4$C$_4$ search has the most degrees of freedom out of any search cell in our program; its winning structure does not appear in the phase diagram.  \label{fig:hartke}}
\end{figure}

We carry out ground-state structure searches using the evolutionary crystal structure prediction software XtalOpt r8.0 \citep{lon11}, together with GULP optimization.\footnote{
It is convenient to reinterpret XtalOpt and GULP's internally-consistent (eV, \AA, GPa) unit system as ($10^d$ eV, $10^{-d}$ \AA, $10^{4d}$ GPa), so that issues with numerical limits can be avoided.  These codes were, of course, originally intended for Earth-condition materials!  Useful choices of the integer $d$ include 2, 3 and 4.  In this scheme, the relativity parameter appearing in Equations \ref{tau0} \& \ref{k0inverse} becomes $x=1.1946484\times10^{d-2}\;n_e^{1/3}$, and the prefactor in Equation \ref{tau0} becomes $m_ec^2/8\pi^2\lambda_e^3=1.1239083\times10^{11-4d}$.}  
It is assumed the ground state is a polycrystalline mixture of stoichiometric compounds.  We don't consider solution (alloy) phases, and we consider only a subset of possible stoichiometries.  For a given ternary system of nuclei $A$-$B$-$C$, a constant pressure search is performed at $P=10^{11}$ and $10^{16}$ GPa for each of the nominally 125 stoichiometries $A_nB_mC_{\ell}$ where $n,m,\ell=0\dots4$. Search cells with small number of particles $n+m+\ell$ are removed from the search program if they are submultiples of larger cells, thus there are 98 searches per ternary system, per pressure. Each of these searches is run out to at least 480 optimized, genetically-operated-on structures, except in the case of single-component searches, which are run out to at least 80 optimized structures (the first 20 seed structures are randomly generated).  Lattice sums are done in real space with cutoff $\approx20k_0^{-1}$ and are expected to be converged to 7--8 digits. This level of accuracy is important, as we find enthalpies of competing structures can be the same out to 6 digits. Default XtalOpt search parameters are used throughout, and following the suggestions put forth in the XtalOpt implementation paper \citep{lon11}, we benchmark the search parameters by constructing Hartke plots for several relativistic screened-Coulomb systems (see Figure \ref{fig:hartke}).  Hartke plots gauge the performance of a genetic search and help establish a stopping criterion.  
Our choice of search duration, previously mentioned, is in part motivated by the ``Hartke lifetimes" found in these tests.  

The lowest enthalpy structure found in each search is included in a bulk phase stability calculation, using Thermo-Calc software \citep{and02}.  For a given set of $N_C+2$ state variables ($N_C$ being the number of components) Thermo-Calc finds the global minimum Gibbs free energy which lies on a plane tangent to the available phases' Gibbs energy surfaces.  A phase diagram representable as a 2d plot is then constructed from the set of tangent planes found by varying any two of the state variables.  (For a pedagogical reference to phase diagrams, see the book by \citet{hil07}).  Since all the phases considered in this work are stoichiometric crystal structures, there is a simplification in that the phases' Gibbs energy surfaces are themselves points.  Consequently, all phase regions in a phase diagram obtained by the above procedure must be 3-phase regions.  If a structure appears in the equilibrium phase diagram at either $P=10^{11}$ or $10^{16}$ GPa, it is re-optimized at intermediate pressure decades to obtain the pressure dependence of the phase diagram.  It is possible, though unlikely, that there are phases stable only over a narrow band of pressure which are missed by this procedure (the full search scheme described in the last paragraph was performed for the C-O-Fe system at several intermediate pressures and found no such ``missed'' phases, lending support to this approach).

Five ternary systems of nuclei were selected for study: He-C-O, C-O-Ne, C-O-Fe, O-Fe-Se, and Fe-As-Se.  (The single exception to the search program described above has to do with He-C-O: our real space method converges much more slowly in this system due to the larger $k_0^{-1}/r_s$, so the full 98 searches are carried out only at $P=10^{11}$ GPa).  The first two ternary systems are relevant to WDs, likely including those which are type Ia supernovae (SNIa) progenitors \citep{she14}.  The third may also be relevant to WDs having undergone a failed-detonation SNIa \citep{jor12}.  The rationale for choosing the remaining two is that these particular nuclei are representative and/or prevalent among the \citet{gup07} abundances near neutron drip. Incorporating a full list of abundances ($\approx\,$17 species) would be intractable for the type of calculation we have described. Moreover, we can take a lesson from earth-condition crystals, which typically have only 1, 2, or 3 elements (sometimes 4).  This appears to be due to general properties of crystal stability related to phase separation of complex unit cells: Basically, once a structure reaches a sufficient level of complexity that it can accommodate the special geometrical characteristics of its constituent atoms, it is disadvantageous to make the unit cell any larger (in the sense of adding more atoms), since that just reduces the amount of favorable repetition possible with a given number of atoms.  Exploring more ternary combinations is also likely to give diminishing returns in the prediction of new structures.  Roughly speaking, with a smooth and spherically-symmetric interaction such as the Yukawa potential, there are only four qualitatively different ternary combinations: one big $Z$ and two small $Z$s, two big and one small, all three mismatched, and all three similar.  As long as the screening length regimes (characterized by $k_0^{-1}/r_s$) are not too different, one expects to see a continuity of structures formed by systems having similar relative $Z$s (or perhaps squared $Z$s), since there are no atom shell effects that come into play.  Another reason for studying the specific ternary systems mentioned above is that they cover at least three (arguably all four) of the qualitatively different combinations.

\section{bulk phase diagram results}

%
%
\begin{figure*}[t]
\vspace{\baselineskip}
\centering
\includegraphics[width=0.95\textwidth]{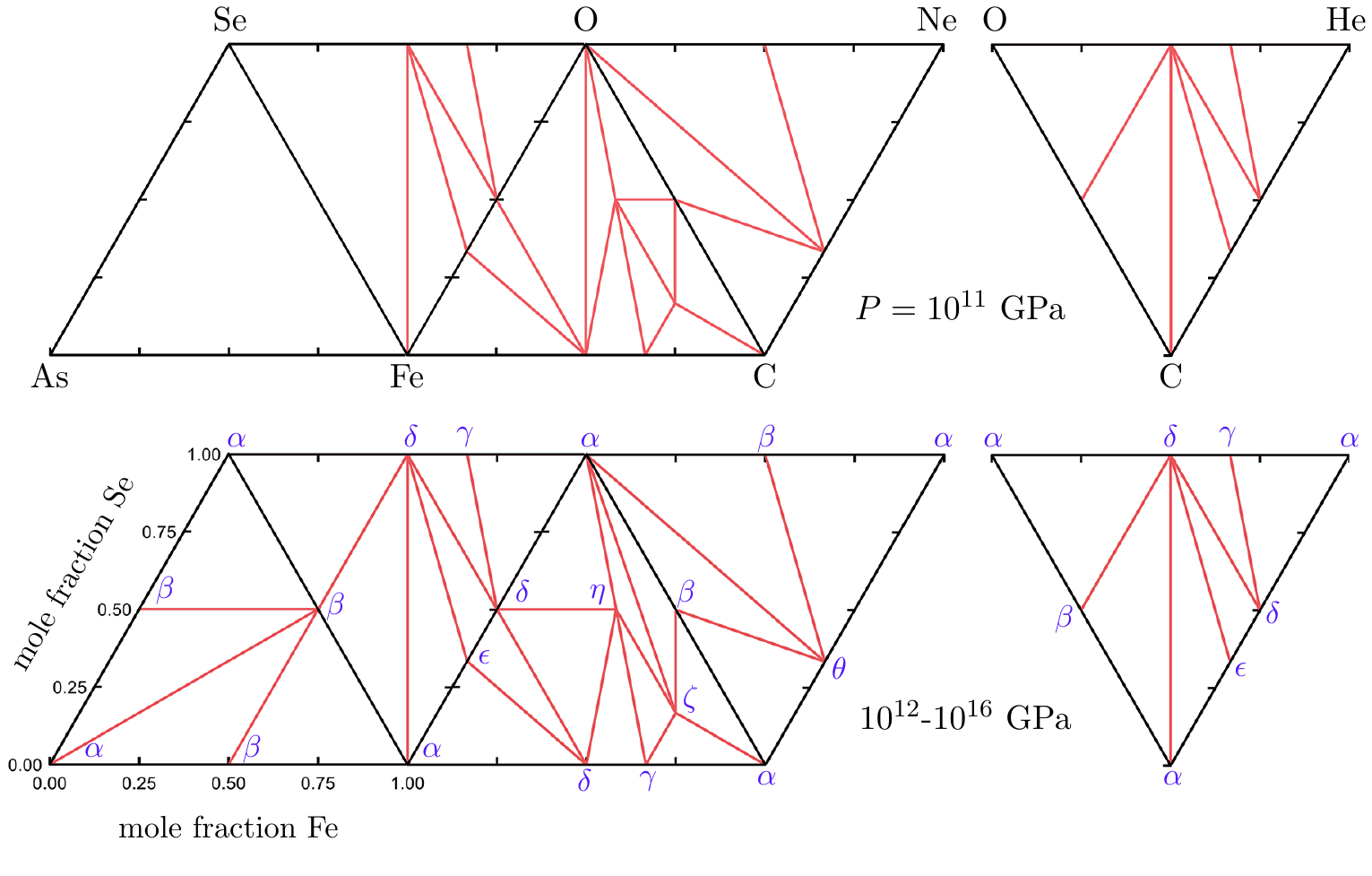}
\caption[width=1.0\textwidth]{$T=0$ bulk phase diagrams for relativistic screened-Coulomb systems.  Each pair of vertically-aligned diagrams corresponds to a specific ternary system of nuclei, while the two rows give the pressure dependence.  Across the five ternary systems investigated, no pressure dependence was found in the range $10^{12}$--$10^{16}$ GPa, despite the fact that the screening length $k_0^{-1}$ varies considerably over this range (from 1.36$r_s$ to 1.81$r_s$ for bcc Fe).  Following from our assumptions described in the main text, the microstructure in a given triangular region is a polycrystalline mixture of stoichiometric compounds (phases).  All distinct phases are labeled with Greek letters and explained in Table \ref{tab:structures}.
\vspace{\baselineskip}
\label{fig:ternary}}
\end{figure*}
While pressure-invariance of the $T=0$ phase diagram was anticipated in the extreme relativistic limit, it comes as some surprise that the pressure-independence persists well below this limit, nearly to the threshold for full pressure ionization.  Figure \ref{fig:ternary} shows that for all five ternary systems studied, no $P$-induced phase transitions were found above $10^{12}$ GPa.  For bcc Fe, this pressure corresponds to density $6.18\times10^5$ g/cc and screening length $k_0^{-1}/r_s=1.36$, or about 75 percent of the saturation value.  In general, screening length on the order of the lattice spacing appears to be a requisite for $P$-driven phase transitions.  Further supporting this conclusion is the fact that the He-C-O and C-O-Ne systems don't undergo any pressure-induced transitions in the range $10^{11}$--$10^{16}$ GPa;  within that range, screening lengths in these small $Z$ systems are significantly larger than one lattice spacing.

Another finding is that combinations of nuclei with significantly mismatched $Z$s are much more conducive to efficient multicomponent packings than are systems where the $Z$s are fairly similar.  For example, the Fe-As-Se system has an extremely simple low-pressure phase diagram: at any composition, the microstructure consists simply of phase-separated bcc crystallites.  Multicomponent phases appear at high pressure, but they have the simple cesium chloride structure.  In contrast, the C-O-Fe phase diagram is quite rich.  Combining one large $Z$ and two smaller $Z$s results in a variety of binary and ternary crystal structures (enumerated in Table \ref{tab:structures}), all of which are more efficient (have a higher packing fraction) than phase-separated bcc lattices.  

Both He-C-O and O-Fe-Se systems (two large, one small) feature all the same phases as C-O-Fe, except for the two ternary compounds which don't appear.  While a continuity of structures appearing between these systems was anticipated, it is striking that at high pressures the two phase diagrams are identical. Close similarity is also noted between the Fe-As-Se and C-O-Ne systems which both consist of three similar $Z$s. The outlier in this comparison is the nontrivial C-Ne binary structure, described in Table \ref{tab:structures}.  These observations are consistent with the idea that it is the combination of relative $Z$s, and not of absolute $Z$s, that is important in determining the high-pressure phase diagram. 
\begin{table*}[t] 
\centering
\caption{Selected compounds appearing in the C-O-Fe and C-O-Ne bulk phase diagrams, as indicated in Figure \ref{fig:ternary}.  All numerical values given here correspond to $P=10^{16}$ GPa.  Relative proton density is a measure of geometrical packing efficiency; relative baryon density includes the nongeometrical effect of neutron fractions.  The reference phase for these relative densities is $\alpha$-Fe, except in the case of $\theta$-Ne$_2$C$_4$, for which the reference phase is $\alpha$-Ne.  Renderings have grey C, red O, green Fe, and violet Ne with sphere volume proportional to the nuclear charge $Z$.  In the $\delta$ and $\epsilon$ renderings, bonds indicate Fe-Fe first nearest neighbors.  If the space group is listed instead of a specific crystal structure, the unit formula gives the composition of the search cell in which the structure was found, not necessarily that of the primitive cell. pdb files of the structures (for all compositional instances) are included as supplementary materials in the online version.} \label{tab:structures}
  \begin{tabular}{p{1.0cm}p{1.4cm}p{3.2cm}lllp{4.5cm}}
      \hline  &  & crystallographic & relative & relative & density relative \\
      phase & unit & space group & proton & baryon & to bulk, phase- & views along (or slightly oblique to)\\
      label  & formula & or structure & density & density\footnote{Using $^{12}$C, $^{16}$O, $^{20}$Ne, and $^{56}$Fe} & separated bcc\footnote{Defined as the sum of cell volumes after phase-separation into bulk bcc phases, divided by the original cell volume\\} & some high-symmetry directions \\ \hline
      $\alpha$ & Fe & bcc & 1 & 1 & 1 & \\
      $\alpha$ & O  & bcc & 0.982$\;\;\;\;\;$  & 0.912$\;\;\;\;\;$ & 1 & \\
      $\alpha$ & C  & bcc & 0.980 & 0.910 & 1 & \\
      $\beta$ & OC & cesium-chloride & 0.981 & 0.911 & 1.000001 & \\
      $\gamma$ & FeC$_2$ & magnesium-diboride & 0.994 & 0.971 & 1.000061 & \\
      $\delta$ & Fe$_4$O$_4$ & Cmcm (orthorhombic) & 0.996 & 0.979 & 1.000040 & similar to $\delta$-Fe$_4$C$_4$, see below \\
      $\delta$ & Fe$_4$C$_4$ & Cmcm (orthorhombic) & 0.996 & 0.983 & 1.000056 & \parbox[c]{1em}{
       	\vspace{6pt}\includegraphics[scale=0.11]{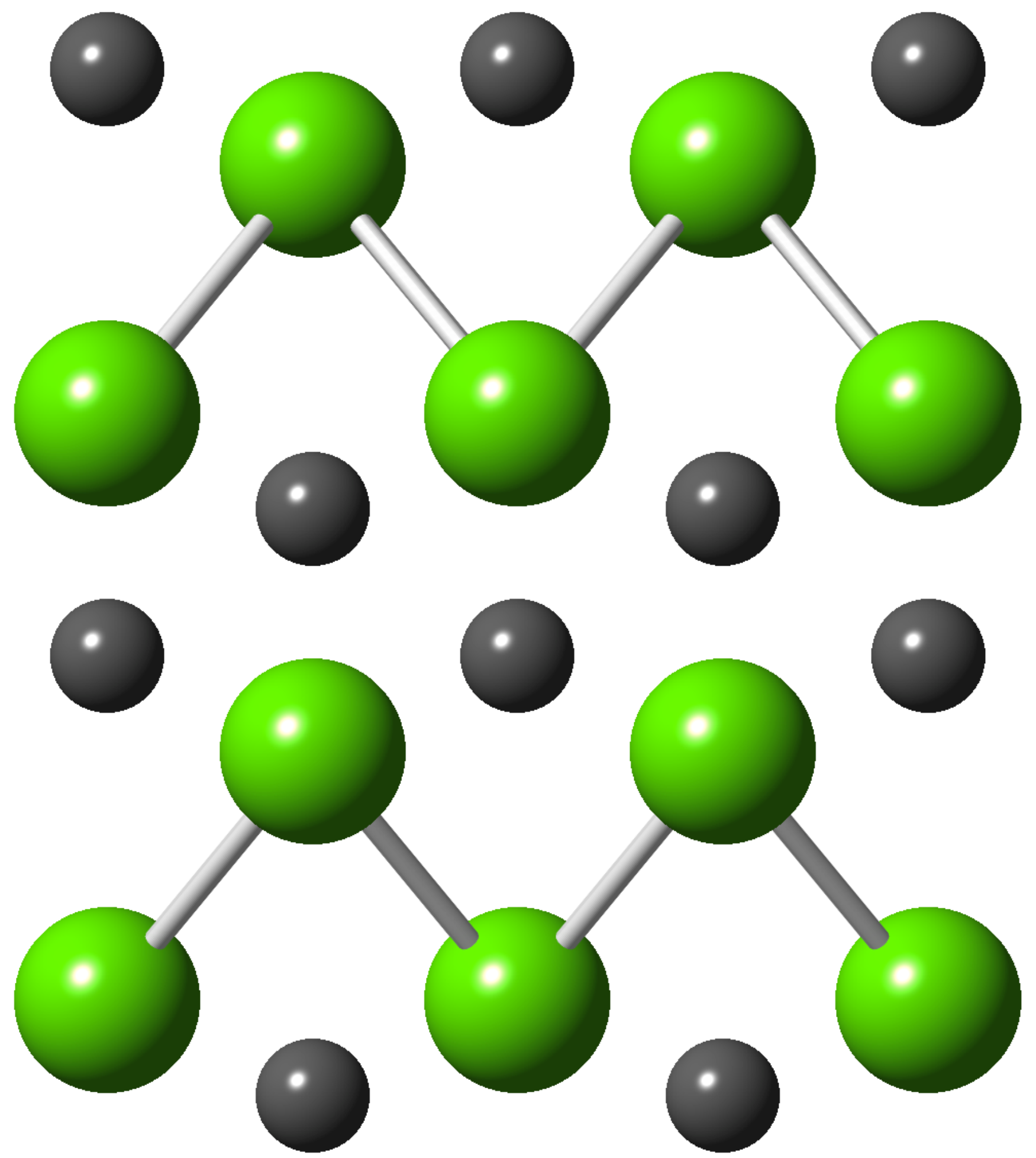}$\;\;\;\;\;\;$\includegraphics[scale=0.11]{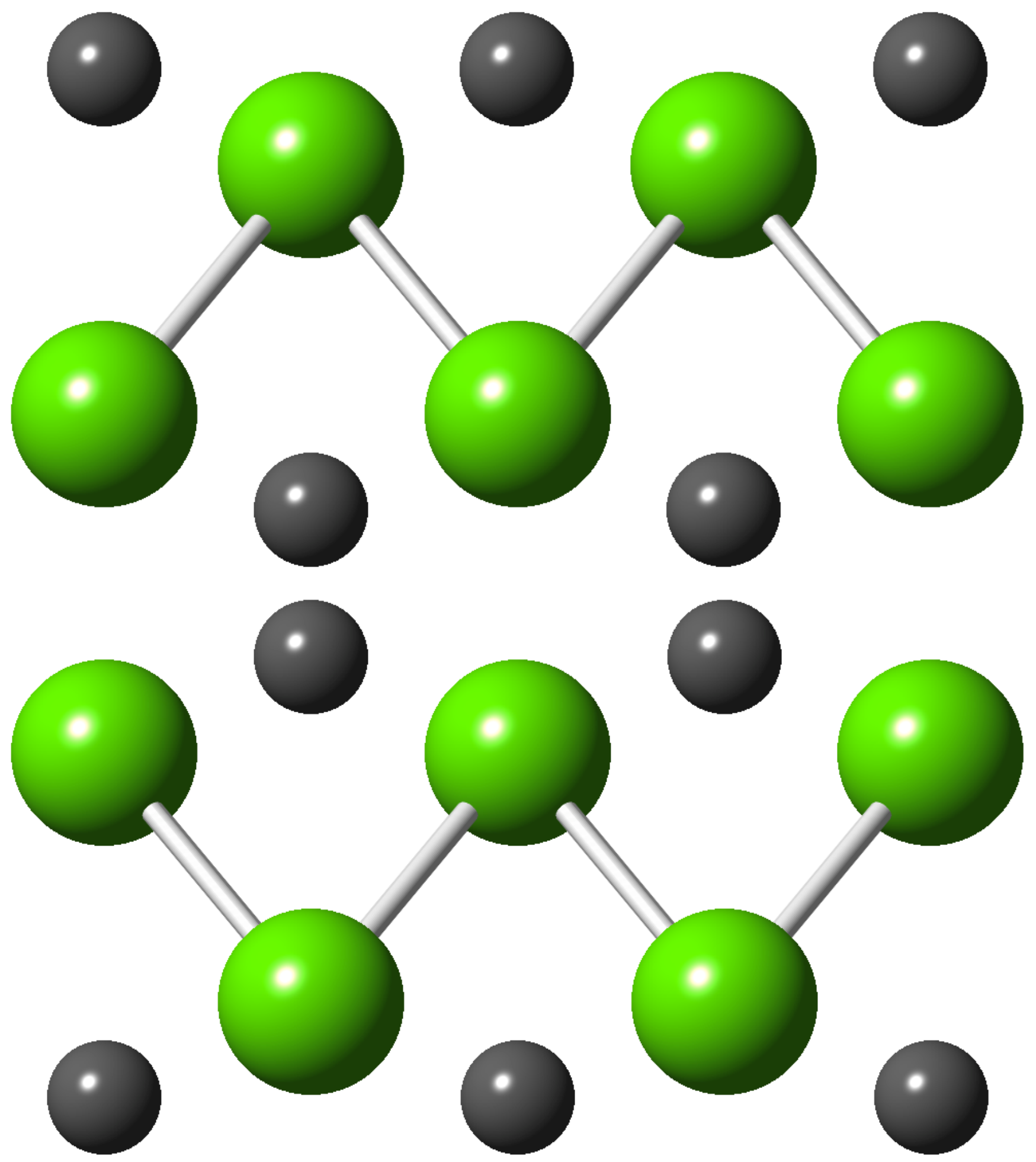}} \\
      $\epsilon$ & Fe$_4$O$_2$ & I4/mcm (tetragonal) & 0.998 & 0.988 & 1.000024 & \parbox[c]{1em}{
	\vspace{6pt}\includegraphics[scale=0.13]{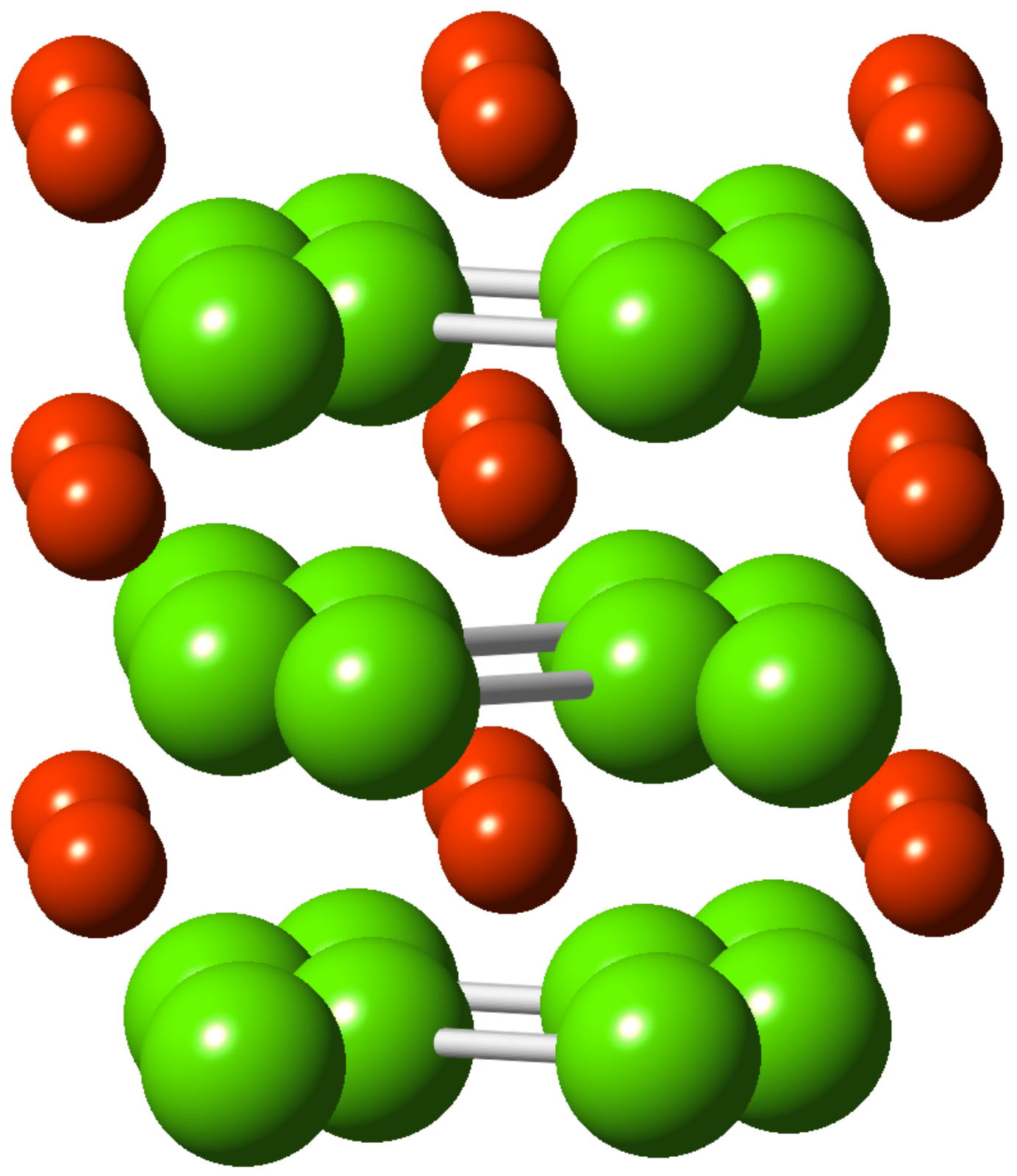}$\;\;\;\;$\includegraphics[scale=0.15]{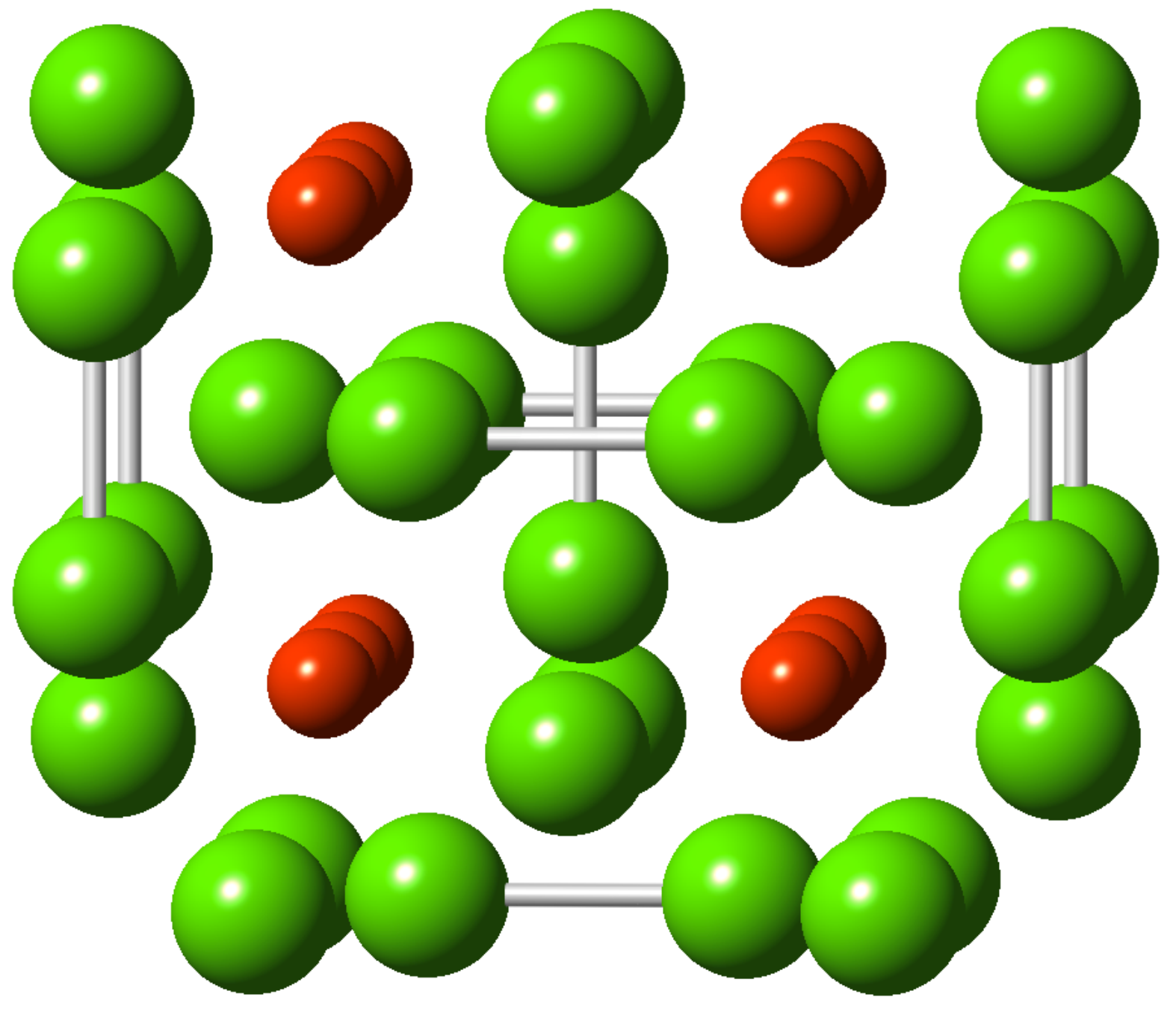}} \\
      $\zeta$ & FeOC$_4$ & P6/mmm (hexagonal) & 0.989 & 0.950 & 1.000044 & \parbox[c]{1em}{
	\vspace{6pt}\includegraphics[scale=0.16]{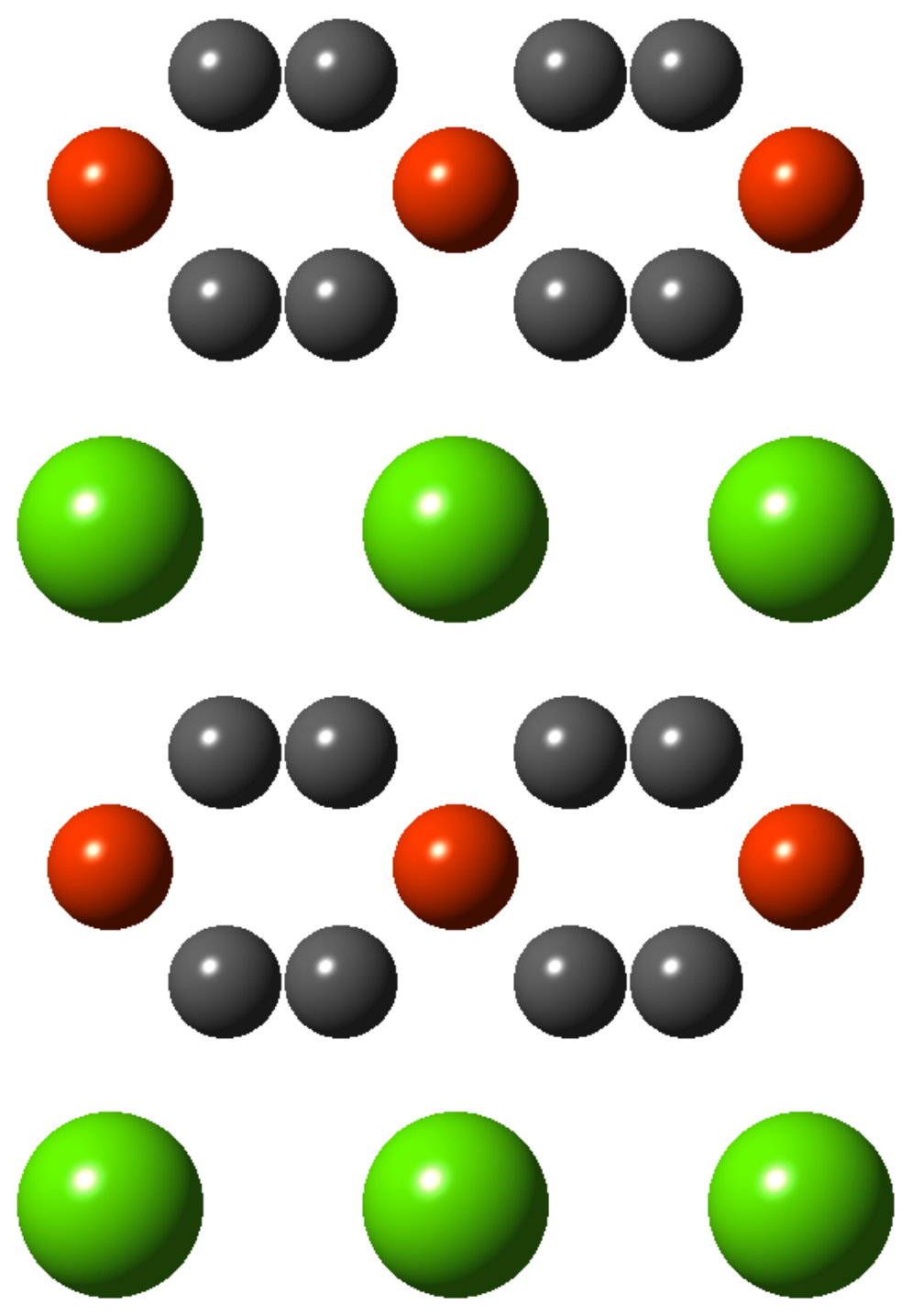}$\;\;\;$\includegraphics[scale=0.14]{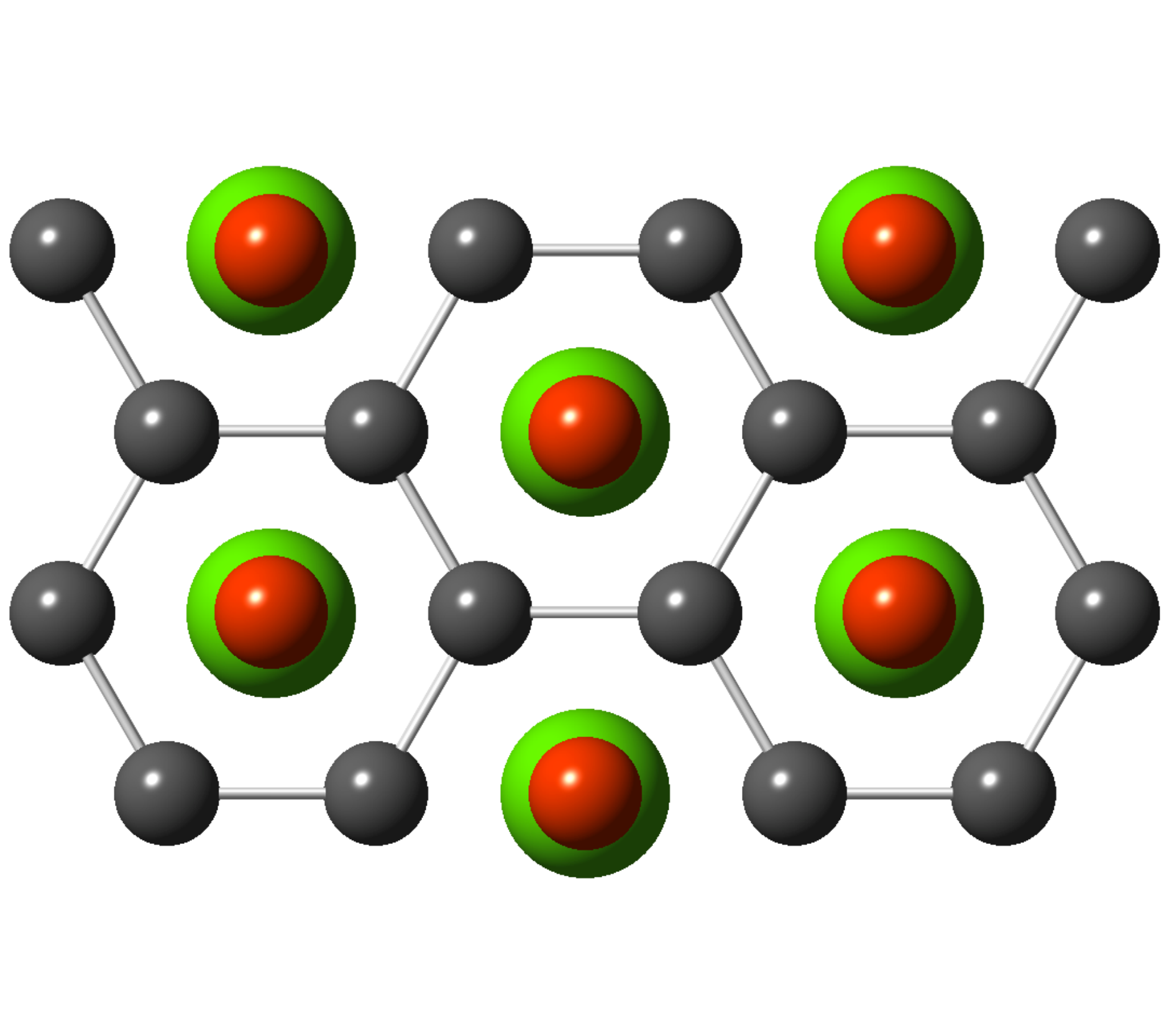}} \\
      $\eta$ & FeO$_3$C$_2$ & P6/mmm (hexagonal) & 0.989 & 0.948 & 1.000039 & \parbox[c]{1em}{
	\vspace{6pt}\includegraphics[scale=0.19]{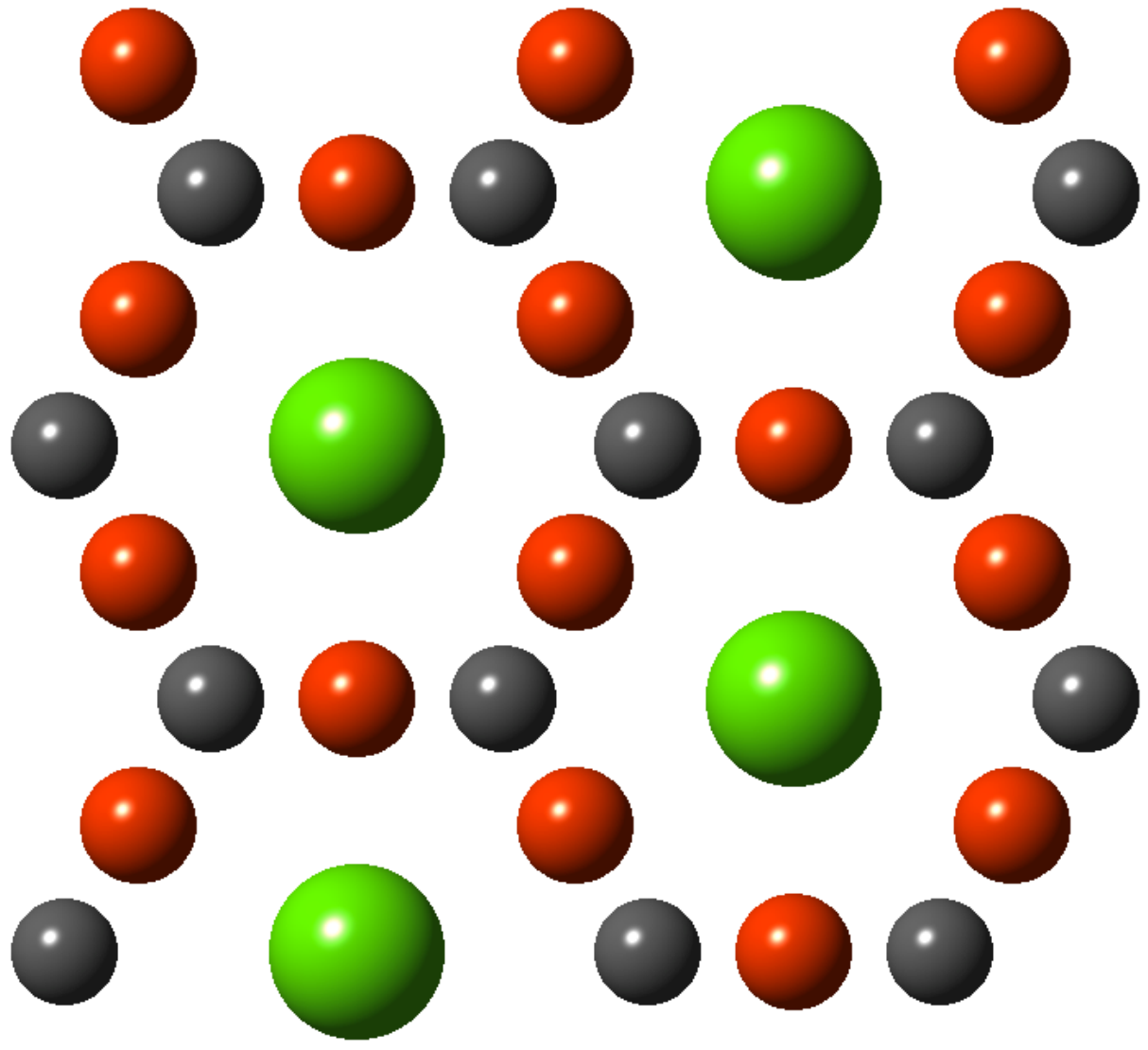}$\;\;\;\;\;$\includegraphics[scale=0.20]{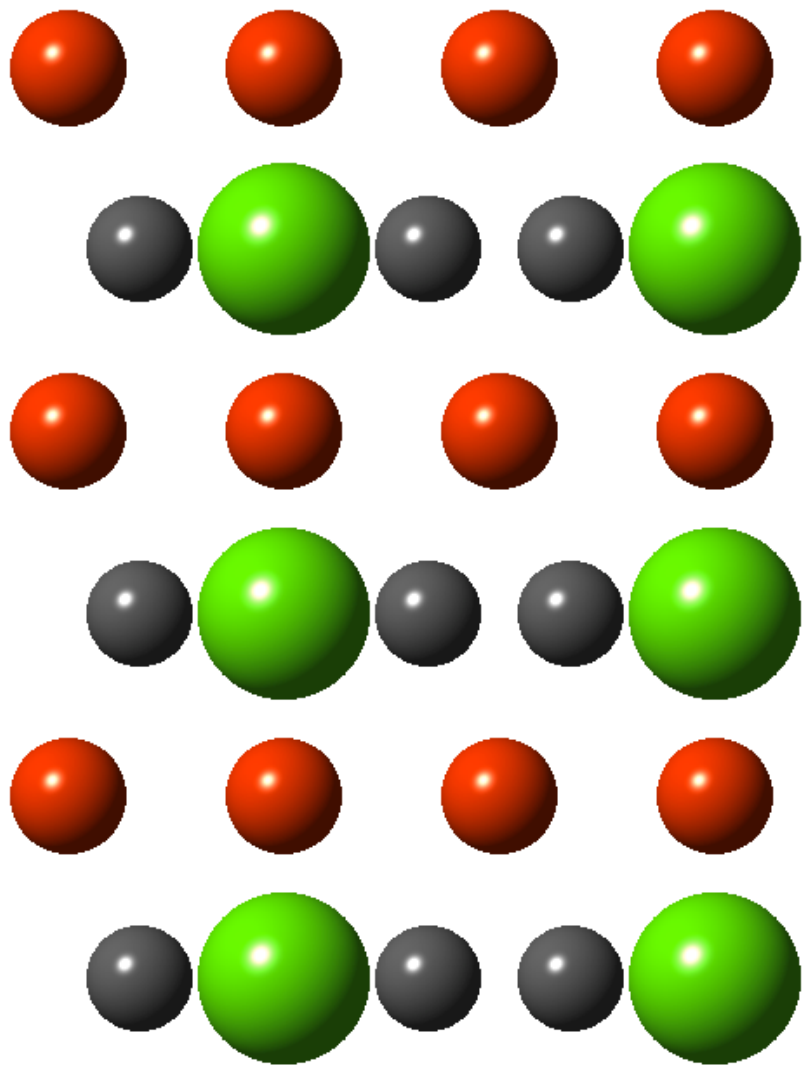}} \\
	$\theta$ & Ne$_2$C$_4$ & Fd-3m (cubic) & 0.997 & 0.997 & 1.000021 & \parbox[c]{1em}{
	\vspace{6pt}\includegraphics[scale=0.2]{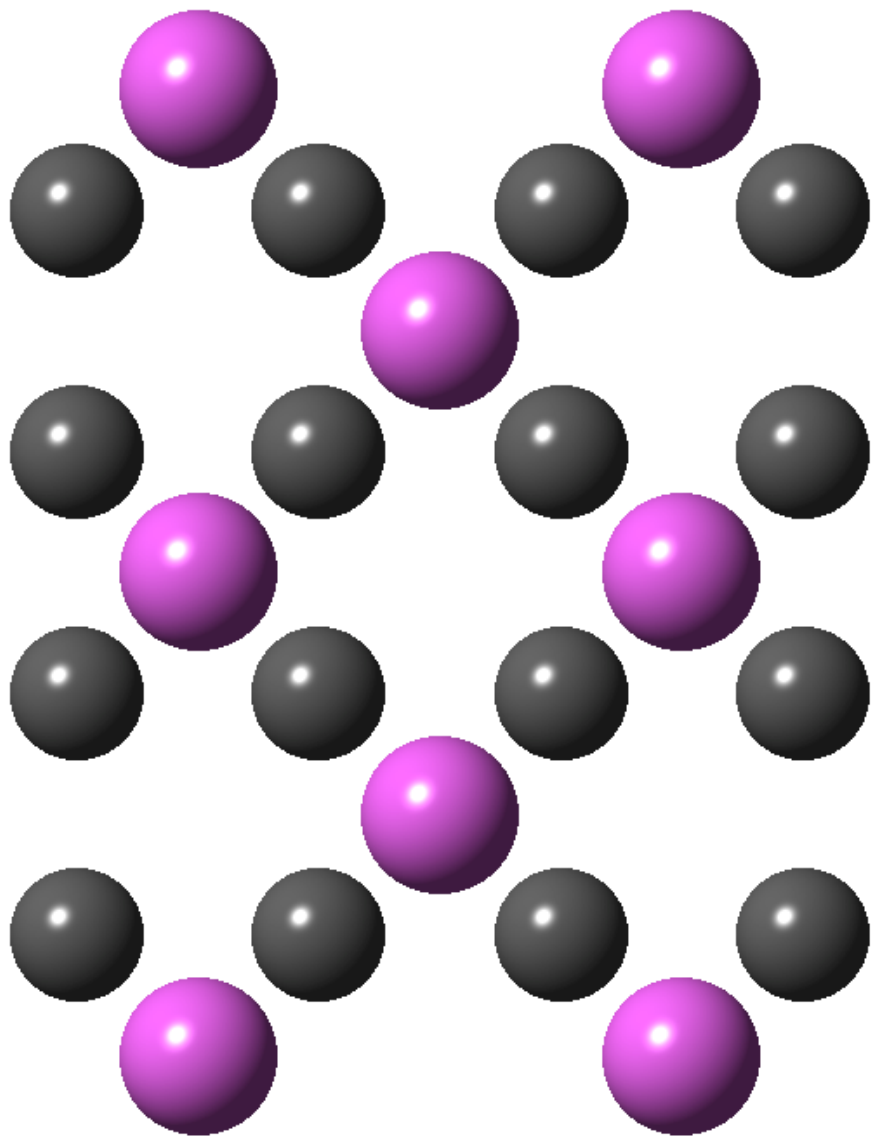}$\;\;\;\;\;$\includegraphics[scale=0.15]{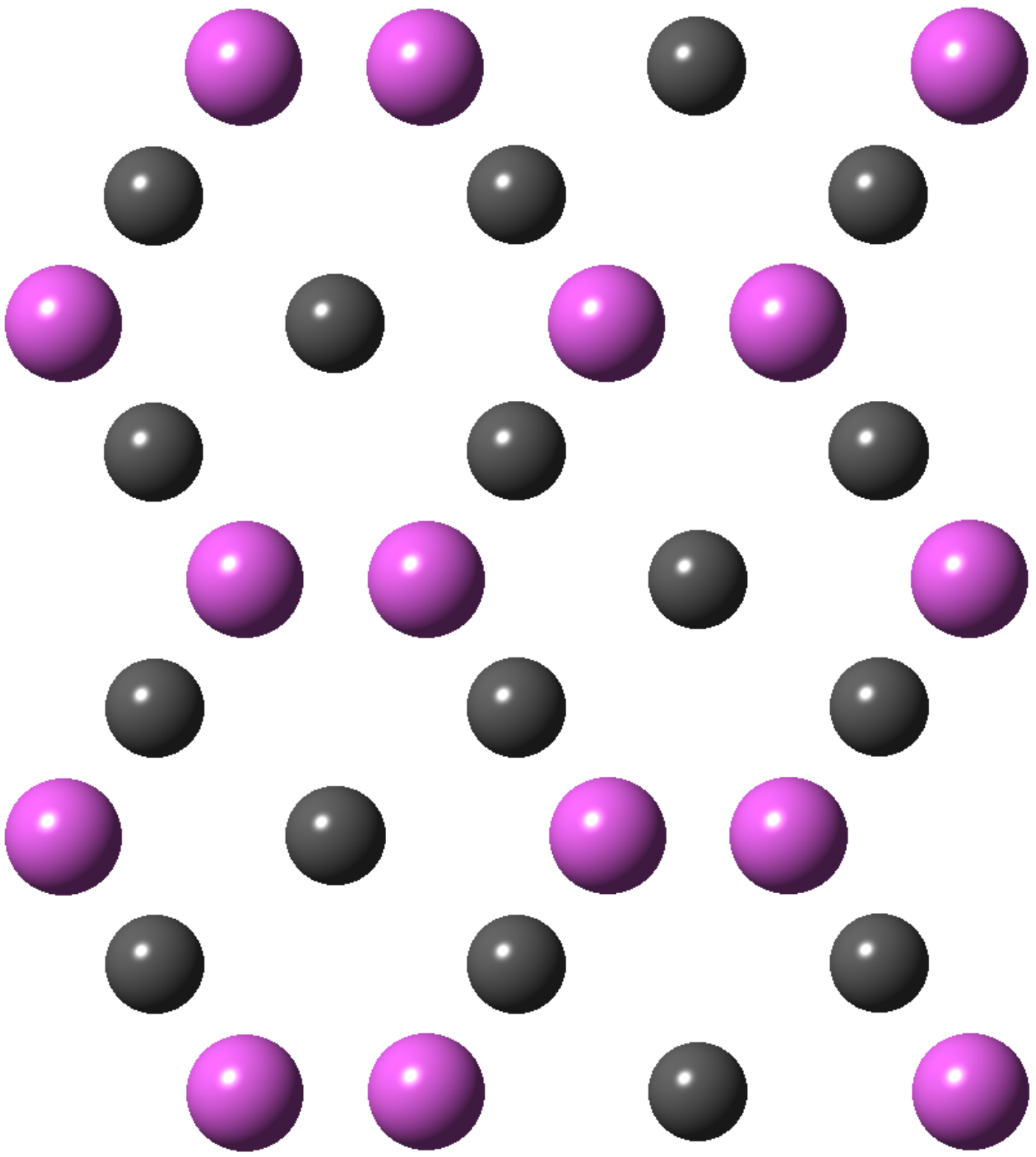}} \\
      \hline
  \end{tabular}
\end{table*}

In the pressure and screening length regimes appropriate to this work (while $P$ ranges from $10^{11}$ to $10^{16}$ GPa, $k_0^{-1}/r_s$ ranges from 1.13 to 1.81 for bcc Fe), there is a competition between close packing and next nearest neighbor interactions, which the closest packed structures tend not to win.  This is exemplified by bcc's favorability over fcc, and the fact that only one of the equilibrium phases found (magnesium diboride structure) also appears in the phase diagram of densest binary sphere packings \citep{hop11,hop12}.  The simplest multicomponent crystals have structures that are also assumed by some ionic compounds under low pressure conditions, which may reflect the fact that ionic solids have a simple close-shell electronic structure (ionic solids also have strong +/$-$ Coulomb interactions that are missing here). When a pair of $Z$s are not too dissimilar they usually form cesium chloride structure, e.g. OC, NeO, SeFe, SeAs and AsFe.  When they are more dissimilar they tend to form magnesium diboride structure, e.g. OHe$_2$, FeC$_2$ and SeO$_2$.  Magnesium diboride is our first encounter with sub-cubic symmetry, which could give rise to transport anisotropy, elastic anisotropy, and other effects such as a magnetic field coupling to the structure orientation.  A quite prevalent but more complicated orthorhombic structure occurs at chemical compositions O$_4$He$_4$, C$_4$He$_4$, Fe$_4$O$_4$, Fe$_4$C$_4$ and Se$_4$O$_4$, where these different instances can be interconverted by small adjustments of bond lengths and angles. A tetragonal structure occurs at compositions C$_4$He$_2$ and Fe$_4$O$_2$; this is the second-highest density structure in the C-O-Fe system and could potentially drive oxygen to greater depths than it would otherwise go.  The C-O-Fe system also features two ternary structures FeOC$_4$ and FeO$_3$C$_2$, both with hexagonal symmetry.  FeOC$_4$ can be viewed as magnesium diboride structure, with the triangular magnesium planes alternating between Fe and O compositions. FeO$_3$C$_2$ consists of alternating layers of kagome O and honeycomb C, with Fe at the holes in the honeycomb layers.

A general feature of the the ternary bulk phase diagrams is that coexisting phases have mass density differences, due to a combination of neutron fraction and geometrical packing effects.  These differences can be as large as $\sim\,$10 percent of the total density and will result in stratification of phase domains in the presence of a gravitational field -- the problem to which we now turn.
\pagebreak

\section{equilibrium layering calculation}

Here we give an application of our high pressure crystal chemistry results to white dwarfs at a given fixed overall composition.  The equilibrium phase-layering diagram of a zero temperature WD is computed self-consistently, allowing for arbitrary numbers of components $N_C$ and phases $N_P$ that can be formed from these components.  The problem is decomposed into two parts:  one part is a microscopic phase stability calculation which produces a function $\rho(h)$ where $\rho$ is density and $h$ is enthalpy per unit mass, the other is a simple stellar structure calculation which determines the pressure-radius dependence $P(r)$. We iterate between these two parts. The former is inspired by a technique used among chemical engineers to study species segregation in oil reservoirs, cf. \citet{esp00}.

In the following, we will make use of the virial theorem for the gravitational potential energy $W$ of a WD, given by 
\begin{equation}
W = -3 \int_0^R P\,4\pi r^2 dr. \label{virial}
\end{equation}
We begin by discretizing the star into $N_L$ onion layers of uniform thickness $\Delta r=R/N_L$.  If $\Delta r$ is small compared to the scale height of pressure $H_P=-dr/d\log P$, the $i^{th}$ layer may be treated as a bulk equilibrium system at constant pressure $P_i$, and one may work with the modified Helmholtz free energy
\begin{equation}
F^* = \sum_{i=1}^{N_L} \Big[ -4P_iV_i + \sum_{\alpha=1}^{N_P} n_{\alpha i}\,\mu_{\alpha}(T,P_i) \Big]. \label{helmholtz}
\end{equation}
For each term in the sum over layers, $-3P_iV_i$ comes from the discrete version of Equation \ref{virial}, and another $-P_iV_i$ cancels the corresponding quantity in the Gibbs free energy of the layer, $\sum_{\alpha}n_{\alpha i}\,\mu_{\alpha}(T,P_i)$. Here $n_{\alpha i}$ is the (unknown) molar amount of phase $\alpha$ present in layer $i$, and $\mu_{\alpha}$ is the bulk chemical potential of phase $\alpha$.  (The phase index $\alpha$ is not to be confused with the bcc structure, as in Table \ref{tab:structures}).  We have thus avoided the complication of introducing a gravitational term into the chemical potentials, including it instead at the level of the layers.  This comes at the cost of supplying a pressure function $P(r)$ consistent with hydrostatic equilibrium, implicit in Equations \ref{virial} \& \ref{helmholtz}.  Let's assume that we have such a pressure function.  (For an initial guess, we will take $P(r)$ from a $n=3$ polytrope.)  Now fix a set of layer pressures $P_i=P(i\Delta r)$.  The problem of minimizing $F^*$ has been reduced to the problem of minimizing the linear objective function $\sum_i\sum_{\alpha} n_{\alpha i}\,\mu_{\alpha}(T,P_i)$, subject to $2N_L+N_C-1$ constraints
\begin{eqnarray}
&&1 = \frac{1}{V_i}\sum_{\alpha} \frac{n_{\alpha i}\,m_{\alpha}}{\rho_{\alpha}(T,P_i)} \;\;\;\;\textrm{for }i=1\dots N_L, \\
&&0 = \sum_{i,\alpha} [(1-x_A)s_{A\alpha} - x_As_{B\alpha} - x_As_{C\alpha} ]n_{\alpha i},\nonumber\\
&& \textrm{etc. for } x_B \dots x_{N_C}, \label{comp} \\
\nonumber\\
&&0 \geq \sum_{\alpha} \Big[\frac{n_{\alpha i+1}}{V_{i+1}} - \frac{n_{\alpha i}}{V_i} \Big] m_{\alpha} \;\;\;\;\textrm{for }i=1\dots N_L\textrm{-1}, \label{BV}
\end{eqnarray}
each of which is also linear in the $n_{\alpha i}$.  The first set of constraints ensures the volume filling fraction is equal to 1 for each layer, where the molar mass $m_{\alpha}$ and density $\rho_{\alpha}(T,P_i)$ are assumed to be known for each phase.  Equations \ref{comp} constrain the global mole fractions $x_A \dots x_{N_C}$, where, for example, $s_{A\alpha}$ specifies the number of $A$-type nuclei per formula unit of the $\alpha$ phase.  The reason for constraining mole fractions rather than component masses is that the latter tends to cause infeasibility problems for the Simplex solver.  Finally, there is a set of inequality constraints which guarantee reality of the inter-layer Brunt-V\"ais\"al\"a frequencies $\omega=\sqrt{-(g/\rho)(d\rho/dr)}$.  Thus, there is built-in stability against convective overturn of adjacent layers, but note that it is still possible to have unstably stratified material \textit{within} a layer.  As noted above, this problem is straightforwardly solved using the Simplex method.  For number of variables $N_LN_P\sim10^3$--$10^4$, we use the high-performace lp\_solve routines.  

So far we have considered the case of an isothermal WD.  For the special case $T=0$, the Simplex solution provides the layer enthalpy per unit mass $h_i=\sum_{\alpha} n_{\alpha i}\,\mu_{\alpha}(0, P_i) / \sum_{\alpha} n_{\alpha i}\,m_{\alpha}$ and layer density $\rho_i=\sum_{\alpha} n_{\alpha i}\,m_{\alpha} / V_i$.  In certain cases, one can interpolate to obtain a smooth function $\rho(h)$, which can be combined with the enthalpy-transformed stellar structure equations \citep{lin92}.\footnote{
An issue can arise near a density discontinuity, where $h_i$ and $\rho_i$ obtained by the above procedure describe a function $h(\rho)$ which is non-monotonic.  The stellar structure calculation cannot then make use of the enthalpy transformation, because the sign of Equation \ref{drdh} is incorrect in the vicinity of the interface.  A density discontinuity occurs as a consequence of mismatched $Z$s (compare bcc C and O in Table \ref{tab:structures}) but is made much more severe when there is also a mismatch in neutron fraction (compare bcc O and Fe in Table \ref{tab:structures}).  Fortunately, for typical white dwarf compositions, the neutron fraction is continuous (or nearly so) across phase boundaries and the issue of non-monotonicity is avoided by choosing a suitably large layer thickness -- on the order of $R/200$ for He-C-O and C-O-Ne compositions.} 
In the nonrelativistic limit, these read
\begin{eqnarray}
\frac{dP}{dh} & = & \rho, \label{dPdh}\\
\frac{dm}{dh} & = & \frac{-4\pi r^4\rho}{Gm}, \\
\frac{dr}{dh} & = & \frac{-r^2}{Gm}.\label{drdh}
\end{eqnarray}
The reason for using the enthalpy transformation is twofold.  First, the total mass $M$, which we were not able to constrain in the Simplex calculation, now enters as a boundary condition.  Second, if we simply used the layer masses $M_i=\sum_{\alpha} n_{\alpha i}\,m_{\alpha}$ along with the discretized equations of mass continuity and hydrostatic equilibrium (the usual stellar structure equations) to update the $P_i$, no information about the microscopic energy scale would carry over.  In other words, we could multiply all the chemical potentials by 2 and get the same $P_i$.  Equations \ref{dPdh}--\ref{drdh} are to be integrated inward from the boundary conditions $P(0)=0$, $m(0)=M$, and $r(0)=R$.  Unfortunately we have no \textit{apriori} knowledge of the radius $R$ that is consistent with $M$ and $\rho(h)$ in the sense of the Oppenheimer-Volkoff map (to use the language of Lindblom).  We therefore have to complete the mapping $\rho(h) \mapsto (M,R)$.  A simple way to accomplish this is by ``aiming" the boundary condition $r(0)$ until the integration yields the physically-correct behavior $dP/dr=0$ as $r\to0$.  Approaching $R$ from below, the solutions are smooth and well-behaved except at $r=0$ due to a singularity in Equation \ref{drdh}.  A change of variable $u=r^2$ removes this singularity, but we find no particular advantage to working with the resulting transformed equations.  Approaching $R$ from above generates sign changes and the solutions are generally chaotic.  The qualitatively different behaviors in these two regimes can be exploited to obtain $R$ to arbitrarily high precision.  In practice, we minimize $dP/dr$ at a fixed, small fraction of the starting boundary condition $r(0)$ (but see the next paragraph for discussion of a special case).  In this process of ``completing the map," an updated pressure function $P(r)$ is obtained at no extra cost. Layer pressures are reassigned and input to the Simplex calculation, and the process iterated.  One choice of convergence criterion is that successive iterations produce stellar radii which are the same to within a tolerance of $10^{-6}R_{\odot}$ -- typically this criterion is met within just a few iterations.  Another choice is that radial positions and thicknesses of phase strata (as fractions of $R$) are static to within the resolution set by the number of simulation layers -- typically this occurs after just one iteration.

The main type of numerical error incurred is of the following nature.  In the first iteration, integration of Equations \ref{dPdh}--\ref{drdh} never proceeds past the point for which we have tabulated $h_i$, $\rho_i$ data available to interpolate within.  This is just a consequence of having used the polytrope initial guess.  In subsequent iterations, however, we are sometimes forced to make a choice: carry out the integration past the highest tabulated $h_i$, replacing interpolations with extrapolations, or simply terminate the integration when interpolations become impossible, using the current value of $dP/dr$ in the aiming procedure discussed above.  We choose the second option.  The miminization problem within the aiming procedure is, in these cases, somewhat ill-defined, tending to generate some numerical noise which is expressed in the layering diagram near $r/R=0$.  For this reason we present layering diagrams as they appear after the first iteration, noting that changes to the layering diagram are already imperceptible by the second iteration, save for an increase in the level of this numerical noise. 
%
%
\begin{figure*}[t]
\centering
\includegraphics[scale=0.84]{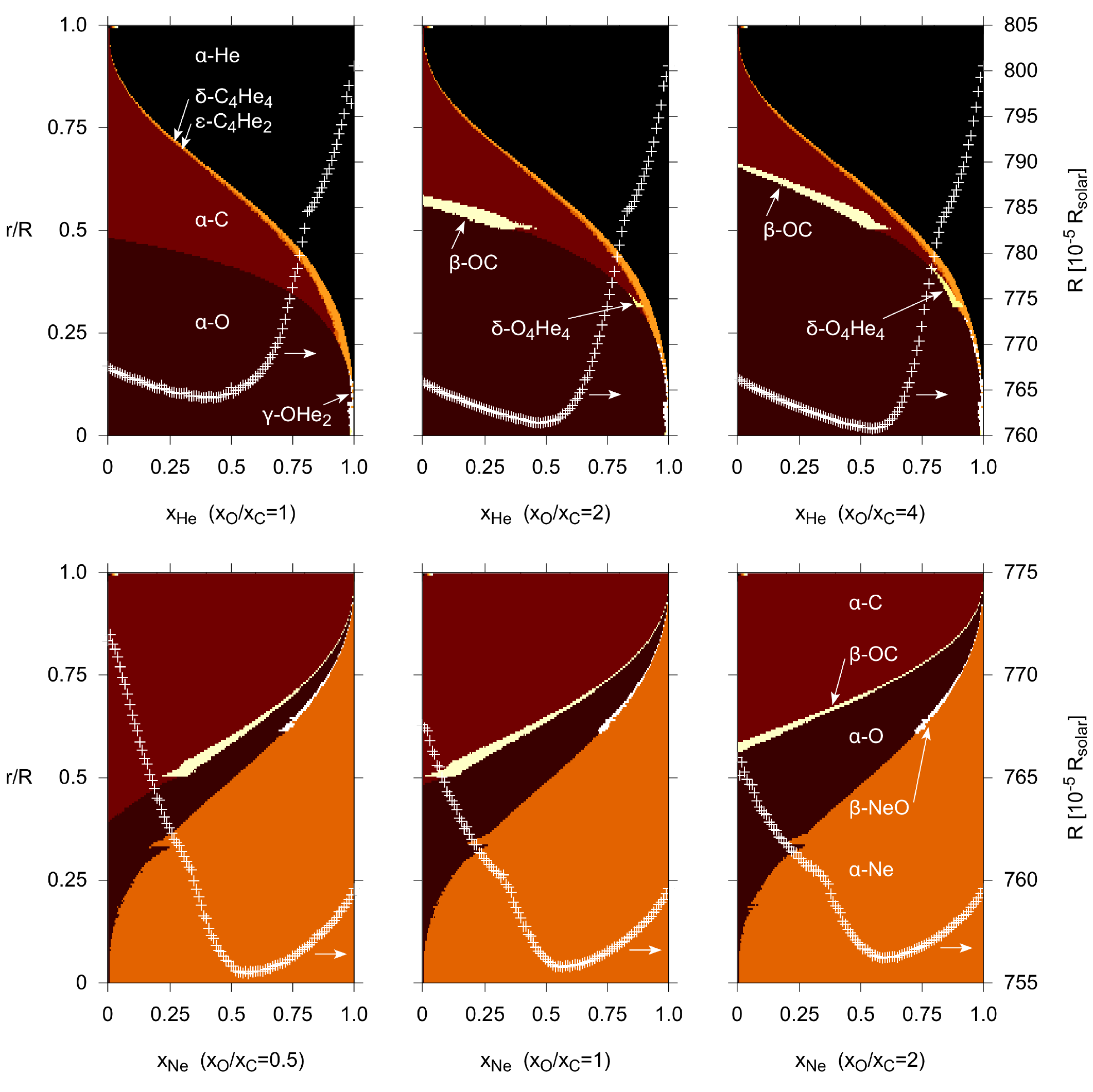}
\caption{  
Equilibrium phase layering diagrams for $1.0M_{\odot}$, He-C-O white dwarfs (top row) and $1.0M_{\odot}$, C-O-Ne white dwarfs (bottom row).  Carbon-oxygen ratio is held at a fixed value in each panel, and mole fraction He or Ne makes up the balance.  The discrete color map indicates the stable phases after one iteration of the procedure described in the main text.  Further iterations produce no perceptible change in the layering diagram, save for an increase in the level of numerical noise.  In cases where a given layer contains a two phase mixture of bcc and a more complicated structure (these are the only types of mixture to occur), the color map indicates the more complicated phase. White crosses give the stellar radius (right y-axis) as a function of composition, also after one iteration of the method.  200 layers were used in the computation of these diagrams.\label{fig:layering}}
\end{figure*}

\section{equilibrium layering results}

The previous section described a method of determining the radial positions and amounts of phase strata in a white dwarf with fixed mass and overall composition, in particular, strata composed of the new multi-component crystal structures.  Using this method, we computed the $T=0$ equilibrium phase layering diagrams and radius-composition dependence of 1.0 solar mass, $^4$He-$^{12}$C-$^{16}$O and $^{12}$C-$^{16}$O-$^{20}$Ne white dwarfs.  For each composition, an initial guess corresponding to the polytrope $P=3.8\times10^{14} \rho^{4/3}$ c.g.s. and stellar radius $R=7.5\times10^{-3}R_{\odot}$ was used, although the method appears to converge to the same result if these starting values are adjusted within reasonable limits. Figure \ref{fig:layering} shows the result of the calculation.  Evidently pure bcc phases make up the majority of the stellar interior, despite the multicomponent structures being more efficiently packed.  Since multicomponent phases tend to show up at interfaces, we refer to them as ``interphases.''   In the He-C-O star, for example, $\delta$-C$_4$He$_4$ and $\epsilon$-C$_4$He$_2$ interphases are formed between $\alpha$-C and $\alpha$-He, while $\beta$-OC appears at the low density part of the C-O boundary.  For compositions near $x_{\footnotesize{\textrm{He}}}=1$, the thinness of the carbon shell allows O-He interphases to form, namely $\gamma$-OHe$_2$ and $\delta$-O$_4$He$_4$.  Compared to sharp bcc-bcc interfaces, interphases offer free energy savings due to optimized crystal packing density, nearest and next-nearest neighbor interactions, etc. arising from the extra compositional degrees of freedom.  Interphase thinness relative to bcc strata can be understood from the gravitational contribution to the free energy having a tendency to ``pull apart" the different Z components of the multicomponent phases.  Competition between these two energy scales apparently causes interphases to become slightly thicker with depth.  Consider, for example, $\delta$-C$_4$He$_4$ in the diagram with $x_{\footnotesize{\textrm{O}}}/x_{\footnotesize{\textrm{C}}}=1$.  At $x_{\footnotesize{\textrm{He}}}=0.5$, only one simulation layer (out of 200) is completely filled with this compound, while an adjacent layer contains a mixture of $\delta$-C$_4$He$_4$ and $\alpha$-He.  This interphase gradually thickens with $x_{\footnotesize{\textrm{He}}}$ and by $x_{\footnotesize{\textrm{He}}}=0.95$, 13 simulation layers are completely filled, another two are partially filled, and $\delta$-C$_4$He$_4$ has squeezed out $\alpha$-C from the layering diagram.

An unexpected but apparently generic feature of the radius-composition curves in Figure \ref{fig:layering} is the existence of a shallow minimum of the WD radius at an impure composition.  Even as the level of numerical noise increases with further iterations, this minimum clearly persists.  The cusps near $x_{\footnotesize{\textrm{He}}}=0.8$ and $x_{\footnotesize{\textrm{Ne}}}=0.3$ may be a numerical effect rather than a physical one, however, as they tend to smooth out upon further iterations.

\section{nonequilibrium layering results}

A simple modification of the equilibrium layering calculation enables a quasi-static settling calculation. If the settling species is $X$, an additional set of linear constraints
\begin{equation}
0 = \sum_{\alpha} s_{X\alpha}n_{\alpha i} \;\;\;\;\textrm{if }i<i_{min},
\end{equation} 
enforces the minimum depth $i_{min}$ at which $X$ can appear.  This minimum allowed depth can then be incrementally stepped down.  We carry out a test of the method by settling $0.09M_{\odot}$ of O on a $0.91M_{\odot}$ He-C white dwarf, and $0.1M_{\odot}$ of Ne on a $0.9M_{\odot}$ C-O white dwarf.  Overall compositions are fixed at $x_{\footnotesize{\textrm{He}}}=0.95$ with $x_{\footnotesize{\textrm{C}}}=x_{\footnotesize{\textrm{O}}}=0.025$, and $x_{\footnotesize{\textrm{Ne}}}=0.07$ with $x_{\footnotesize{\textrm{C}}}=2x_{\footnotesize{\textrm{O}}}=0.62$, respectively; the final settled-out states are given by Figure \ref{fig:layering}.  (Admittedly, these are not particularly realistic settling scenarios but they serve as interesting test cases, forcing the presence of an interface between the highest and lowest $Z$s which otherwise doesn't happen in equilibrium). Note that if the starting value for $i_{min}$ is too near the surface, there is no feasible solution that can accommodate the settling mass.  For this reason we restrict our study to the second half of the settling problem: $i_{min}=N_L/2,\dots,0$.  

\begin{figure}[t]
\centering
\includegraphics[scale=0.62]{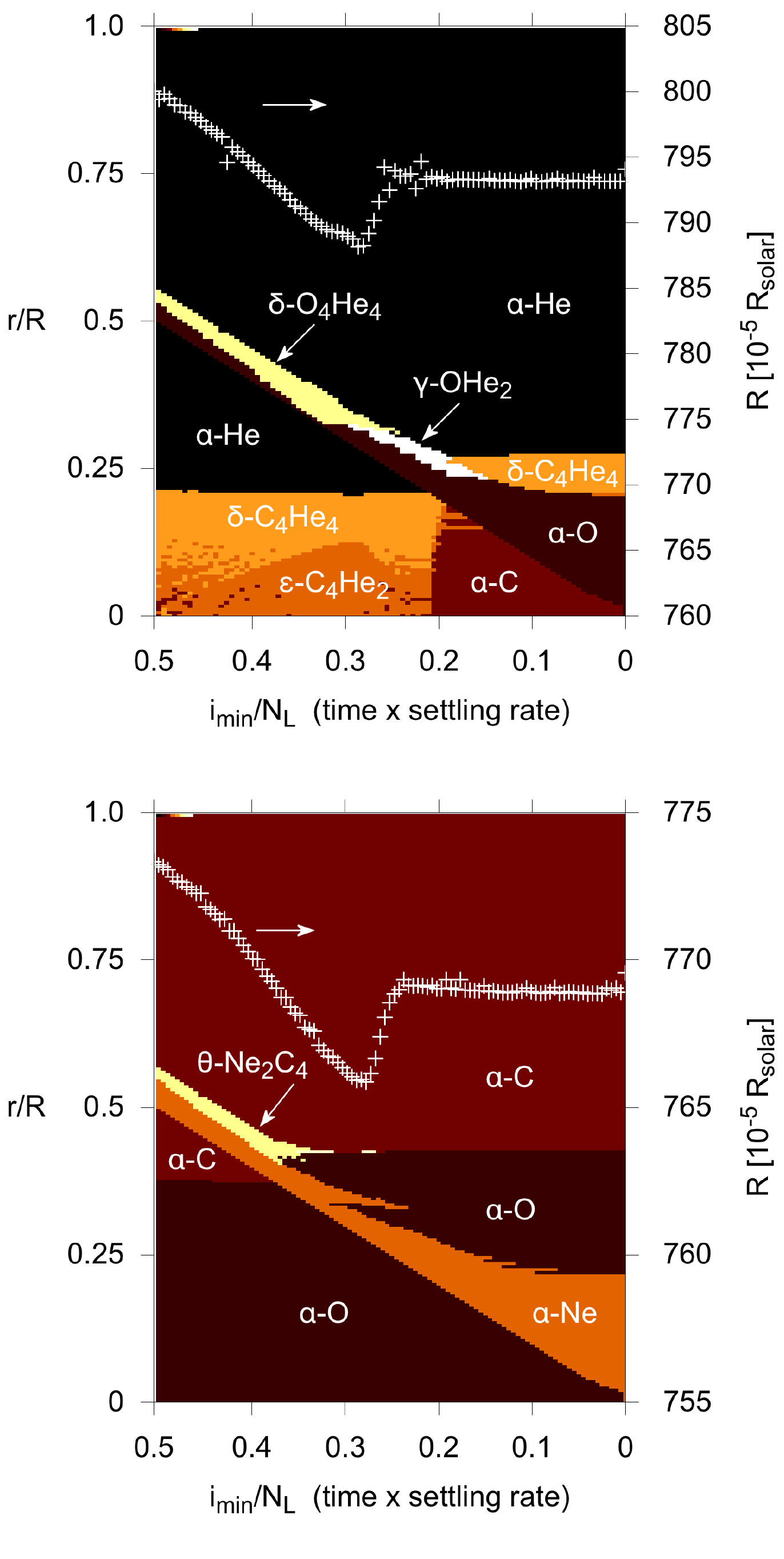}
\caption{  
Quasi-static nonequilibrium phase diagram for settling $0.09M_{\odot}$ of O on a $0.91M_{\odot}$ He-C white dwarf (top) and settling $0.1M_{\odot}$ of Ne on a $0.9M_{\odot}$ C-O white dwarf (bottom).  The $x$-axis can be regarded as time remaining until equilibrium, multiplied by the settling rate.  White crosses give the evolution of the stellar radius during the settling process.  The top(bottom) diagram was computed using 180(200) layers.  Other details of the plots are the same as in Figure \ref{fig:layering}.  
\label{fig:settling}}
\end{figure}

Results of the quasi-static settling calculation are plotted in Figure \ref{fig:settling}.  
In both settling scenarios, the out-of-equilibrium star contains one or more phases that do not appear in the final, equilibrium stacking sequence.  One function of these extra phases is to serve as transient host structures for the settling species:  $\delta$-O$_4$He$_4$ and $\gamma$-OHe$_2$ are hosts for settling oxygen, and $\theta$-Ne$_2$C$_4$ is a host for settling neon. The phase settling diagram for the He-O-C star is fairly non-trivial, with as many as seven distinct strata near $i_{min}/N_L\approx0.25$.  A minimum in the stellar radius appears around this point (as it also does in the O-C-Ne settling calculation) indicating that the lowest enthalpy star is not the most compact star.   This minimum appears to track the size of the $\epsilon$-C$_4$He$_2$ core, possibly also the thickness of the $\delta$-O$_4$He$_4$ interphase.  However, the minimum persists when the calculation is repeated excluding first one and then the other of these phases.  These result hints at the prospects for new phenomena that are enabled by compositional and structural heterogeneity that takes advantage of the additional ``chemical'' degrees of freedom afforded by multinary phases.    

In the first settling calculation, both He and C must eventually find their way through the sinking O-containing layer.  It is interesting that, with the exception of a single point near $i_{min}/N_L=0.3$ where all the oxygen is bound up in $\delta$-O$_4$He$_4$, there is no continuous migration pathway (in the sense of stoichiometric compounds) assuming any deviations from spherical symmetry are sufficiently weak to maintain contiguity of the layering sequence.  Several binary phases are available to provide such a pathway, but the system does not use them to this advantage -- for example, $\beta$-OC is never formed.  Consequently, in the final stages of settling, carbon has to diffuse through an oxygen barrier having thickness $t\sim10^3$ km.  The associated timescale can be estimated as $\tau\sim t^2/D_0$, where $D_0=3\Omega_Pr_s^2/\Gamma^{4/3}$ is the diffusion coefficient for a one-component plasma (see \citet{han75} and more recently, \citet{hug10}).  Here $\Omega_P=(4\pi e^2Zn_e/M)^{1/2}$ is the ion plasma frequency and $\Gamma=Z^2e^2/(r_sk_BT)$ is the Coulomb coupling parameter, which we take to be the melting point value: $\Gamma_m=175$.  Putting in the numbers gives $\tau\sim10^{13}$ yrs.  While this is an oversimplified analysis, it does suggests that compact object phase strata may be far from the equilibrium stacking sequence.  It is particularly intriguing to consider whether strong ``chemical'' deviations from the equilibrium phase stacking sequence could accumulate excess free energy that is eventually liberated in energetic (and thus observable) events. Our analysis here is a first step towards developing a framework towards consideration of such possibilities.

\section{Finite temperature}

Here we give a brief, qualitative discussion of some effects that will become important at finite temperatures;  a detailed analysis is a subject for future work.  At finite $T$, the chemical potentials $\mu_{\alpha}(T,P_i)$ must be modified to account for smearing of the Fermi surface, associated smearing of the electron capture layers which will give an adjustment in composition, phonons, and if $\alpha$ denotes an alloy or solution phase, mixing entropy.  Since the entropic part of these contributions does not enter into the enthalpy, it is not possible to self-consistently include thermal effects in our equilibrium phase layering method, which relies on the enthalpy transformation.  However, thermal effects could be included post hoc.  For example, one could compute the phonon free energy of an interphase crystal such as $\delta$-C$_4$He$_4$ along with that of the equivalent phase-separated $\alpha$-C and $\alpha$-He crystals, add this quantity to the $T=0$ free energy, and predict whether the interphase tends to thicken or thin at finite $T$.  A rough estimate based on phases' bulk moduli ($K=-VdP/dV$) suggests the thermal (phonon) correction to the free energy will tend to increase the stability of the soft outer phase strata relative to the stiff innner strata, likely shifting the interphases slightly towards the stellar center.  There is also a possibility for additional, particularly soft interphases to appear in the stacking sequence, if entropic terms are large enough to affect the phase competition that winnows the phases of Table \ref{tab:structures} down to the phase layering in Figure \ref{fig:layering}.  A phonon calculation would also yield the elastic tensor and help characterize the degree of elastic anisotropy as a function of depth, as well as give a prediction for the relative melting temperatures via the Lindemann parameter (ratio of RMS nucleus displacement to equilibrium lattice spacing).  Finally, we note that the simple mechanical stability criterion used here should be replaced with the Ledoux criterion at finite $T$.  See \citet{rei01} for an application to multicomponent neutron stars.  Again, it is difficult to see how this more sophisticated criterion can be built in to the current method, but it could at least be checked after including thermal effects in the manner described.

\acknowledgments
T.A.E. acknowledges an Academic Computing Fellowship from The Pennsylvania State University.  We thank Julian Gale for providing a GULP patch with the necessary infrastructural changes to handle a $V_c$-dependent screening length, and thank Ben Owen and Steinn Sigurdsson for stimulating discussions.  Teresa Hamill's preliminary investigation of binary crystal packings helped motivate some of the methods outlined in Section 2.
\pagebreak

\end{document}